\documentclass[aps,prd,amsfonts,amsmath,showpacs]{revtex4}
\usepackage[dvips]{graphicx}
\usepackage{bm}

\begin{document}

\title{Point Source Contamination in CMB Non-Gaussianity Analyses}

\author{Daniel Babich}

\email{babich@tapir.caltech.edu}

\affiliation{Theoretical Astrophysics, California Institute of Technology, Pasadena, CA 91125 U.S.A.}

\author{Elena Pierpaoli}

\email{pierpaol@usc.edu}

\affiliation{University of Southern California, Los Angeles, CA 90089 U.S.A.}

\date{\today. To be submitted to Phys. Rev. D.}


\begin{abstract}  	
In this paper we analyze the biasing effect of point sources, either thermal Sunyaev-Zeldovich clusters 
or standard radio sources, on the estimated strength of the non-Gaussianity in the Cosmic Microwave Background (CMB). 
We show that the biggest contribution comes from the cross--correlation of the CMB with the matter density rather 
than from the poisson term which is conventionally assumed in these calculations. For the three year WMAP data, 
we estimate that point sources could produce a non--Gaussian signature equivalent to a bias in $f_{NL}$ of 
$0.35, 0.24, -0.097, -0.13$ in the Ka, Q, V and W bands respectively. The level of bias we find is largely insufficient 
to explain the very high $f_{NL}$ values recently detected by Yadav and Wandelt. 
For Planck, we estimate the point source bispectra to contaminate the $f_{NL}$ estimator with a bias of $1.3, 0.34, -0.25, -0.48$
at $30, 44, 70, 100~{\rm GHz}$ respectively. These results depend on the assumed redshift distribution of the point sources. 
However, given the projected Planck sensitivity of $\Delta f_{NL} \simeq 5$ (95 \% C.L.), a good estimate of point sources' properties 
including their number density and redshift distribution is essential before deriving strong conclusions on primordial 
non--Gaussianity.

\end{abstract}

\maketitle

\section{Introduction}

Recent claims by Yadav and Wandelt \cite{Yadav07} of the detection of strong primordial non-Gaussianity in the three-year 
Wilkinson Microwave Anisotropy Probe (WMAP) data \cite{Spergel07} have the potential to revolutionize our understanding of 
the early universe. These results were also found in the WMAP five-year analysis, although with less statistical 
significance \cite{Komatsu08}. The strength of the non-Gaussianity detected in their analysis is more than two orders of 
magnitude larger than the non-Gaussianity expected in the simplest model of single field, slow roll model of inflation 
\cite{Maldacena03}. This detection, if it stands up to scrutiny, will be the first definitive indication that the simplest 
model of inflation cannot adequately explain all of the current cosmological observations and must be modified in some way. 

In Yadav and Wandelt's analysis \cite{Yadav07}, the detection of the non-Gaussian signal is due to the simultaneous 
reduction in the estimator's error bars and the shift in the central value as smaller angular scale information
was included in the analysis. One obvious concern is that the estimator is contaminated by foreground emission, 
in particular radio points sources and the thermal Sunyaev-Zeldovich effect which become increasingly more important 
on small angular scales. 

In this paper we will analyze the influence of point sources, for the experiment resolution of WMAP both the radio sources 
and SZ clusters effectively act as point sources, on the standard non-Gaussianity estimator.
While it has been claimed that the non-Gaussianity caused by Poisson fluctuations in the number density of radio point
sources can be safely separated from the primordial non-Gaussian signal \cite{Komatsu02}, we will demonstrate
the other forms of point source non-Gaussianity cannot be safely ignored. In addition to the standard forms 
of non-Gaussianity produced by point sources, we will show that cross-correlation between the point source 
power spectrum and the CMB temperature anisotropies or instrument noise will produce non-Gaussianity of a very
similar form as the local model. The local model is the form typically assumed in analyses of primordial 
non-Gaussianity, so these new non-Gaussian contributions will bias the estimator in a fashion that cannot easily 
be corrected.

This paper is organized as follows. In \S \ref{sec:ps} we demonstrate that point sources can produce a 
bispectrum that has the same form as the local model. In \S \ref{sec:bias} we derive the bias induced by 
the various point source bispectra. In \S \ref{sec:conc} we conclude. We use the WMAP three-year cosmological
model \cite{Spergel07} for numerical calculations.

\section{Estimator Bias}\label{sec:bias}
 
While non-Gaussianity generically implies that any higher-order connected correlation function is non-zero,
it is typical to focus on the three-point correlation function, or equivalently the bispectrum, because 
it has the simplest form of all non-Gaussian correlation functions and for weak non-Gaussianity it 
contains the nearly all of the information \cite{Babich05}. The three-point correlation function can be
factored into a component fixed by rotational invariance, which is assumed {\it a prioiri}, and a piece
determined by the underlying mechanism that produced the non-Gaussianity \cite{Komatsu02}. Rotational 
invariance forces the three-point correlation function to be proportional to the Gaunt integral, 
\begin{equation}
\mathcal{G}^{\ell_1 \ell_2 \ell_3}_{m_1 m_2 m_3} = \sqrt{\frac{(2 \ell_1 + 1)(2 \ell_2 + 1)(2 \ell_3 + 1)}{4\pi}}
\left(\begin{array}{ccc} \ell_1 & \ell_2 & \ell_3 \\ m_1 & m_2 & m_3 \end{array}\right)
\left(\begin{array}{ccc} \ell_1 & \ell_2 & \ell_3 \\ 0 & 0 & 0 \end{array}\right),
\end{equation}
and the CMB three-point correlation function can be written as
\begin{equation}
\langle a_{\ell_1 m_1} a_{\ell_2 m_2} a_{\ell_3 m_3} \rangle = \mathcal{G}^{\ell_1 \ell_2 \ell_3}_{m_1 m_2 m_3} 
b_{\ell_1, \ell_2, \ell_3},
\end{equation}
where the reduced bispectrum, $b_{\ell_1, \ell_2, \ell_3}$, contains information about the form of non-Gaussianity. 

The estimators used in CMB non-Gaussianity analyses are optimized for the detection of a signal with a very particular 
form, namely the local model \cite{Babich05}. The local model assumes that the initial curvature perturbations can be written as
\begin{equation}\label{eq:local}
\Phi({\bm x}) = \phi_g({\bm x}) + f_{NL}[\phi^2_g({\bm x}) - \langle \phi^2_g({\bm x}) \rangle ],
\end{equation}
where $\phi_g$ is Gaussian. The non-linear terms in this model lead to the following bispectrum for the initial
curvature perturbations
\begin{equation}
B(k_1,k_2,k_3) = 2f_{NL}[P(k_1)P(k_2) + {\rm cyc.}],
\end{equation}
where $P(k)$ is the power spectrum.
The ordinary linear radiative transfer function are subsequently used to calculate the CMB temperature anisotropies 
from these initial curvature perturbations. The statistical properties of the CMB temperature anisotropies will mirror
the statistical properties of underlying curvature perturbations since we are consider linear radiative transfer. The
levels of non-Gaussianity claimed by Yadav \& Wandelt are significantly larger than the expected non-Gaussianity produced
by non-linear radiative transfer.

However the estimators are sensitive to any bispectrum 
form that might be present in the data, regardless of its origin. In this section we will determine the induced bias 
produced by the various cross-correlation bispectra described in the next section.

The non-Gaussianity estimator can be expressed as \cite{Creminelli06}
\begin{equation}\label{eq:est}
\hat{f}_{NL} = \frac{1}{A} \left[ \sum \mathcal{G}^{\ell_1 \ell_2 \ell_3}_{m_1 m_2 m_3} 
\frac{b_{\ell_1, \ell_2, \ell_3}}{C^T_{\ell_1} C^T_{\ell_2} C^T_{\ell_3}}
a_{\ell_1 m_1} a_{\ell_2 m_2} a_{\ell_3 m_3} \right], 
\end{equation}
where the normalization is
\begin{equation}
A  = \sum \frac{(2 \ell_1 + 1)(2 \ell_2 + 1)(2 \ell_3 + 1)}{4\pi}
\left(\begin{array}{ccc} \ell_1 & \ell_2 & \ell_3 \\ 0 & 0 & 0 \end{array}\right)^2
\frac{b^2_{\ell_1, \ell_2, \ell_3}}{C^T_{\ell_1} C^T_{\ell_2} C^T_{\ell_3}}, 
\end{equation}
here $C^T_{\ell} = C_{\ell} + C^N_{\ell}$ is the sum of the CMB signal and noise. The CMB experimental noise parameters
are described in Table \ref{table:exp_info}.

We are ignoring the additional linear term in the estimator because the contribution of radio point source will not bias 
this piece of the estimator if the signal and noise covariance matrices well represent the real data.

The weight functions used in the estimator are optimized for a bispectrum produced by the local model. The estimator bias 
will be determined by substituting the various forms of the point sources bispectra into the estimator 
\begin{equation}
\Delta f^{\alpha}_{NL} = \frac{1}{A} \sum \frac{(2 \ell_1 + 1)(2 \ell_2 + 1)(2 \ell_3 + 1)}{4\pi}
\left(\begin{array}{ccc} \ell_1 & \ell_2 & \ell_3 \\ 0 & 0 & 0 \end{array}\right)^2
\frac{b_{\ell_1, \ell_2, \ell_3} b^{\alpha}_{\ell_1, \ell_2, \ell_3}}{C^T_{\ell_1} C^T_{\ell_2} C^T_{\ell_3}}.
\end{equation}
Here $b^{\alpha}_{\ell_1, \ell_2, \ell_3}$ is the one of the reduced bispectrum produced by point sources. The CMB bispectrum
produced by the local model is generally negative. The collapsed triangle modes, which have the highest signal-to-noise, are
always negative. So a positive point source bispectrum will cause a negative estimator bias.
We will now discuss the possible bispectrum terms. 

\section{Point Source Bispectra} \label{sec:ps}

The observed signal is the sum of the primordial and secondary temperature anisotropies, foreground emission and 
instrument noise. While the secondary anisotropies and extra-galactic foregrounds, which maybe quite non-Gaussian,
are important on small angular scales, the signal on the large angular scales relevant for WMAP is dominated by
primary anisotropies. Thus it is assumed that any measured non-Gaussianity by WMAP is primordial in nature. We will
argue that cross-correlations between some of these signals may induce bispectra on large angular scales. 

Moreover these bispectra may have similar forms as the local model if the signal power spectrum becomes spatial 
inhomogeneous in manner that then correlates with a second component in the observed signal. For example, the matter 
overdensity will bias the local radio point source power spectrum and it will be correlated with the CMB temperature 
anisotropies produced via the ISW effect. This will produce a bispectrum similar in form to the local model. 

This is not a coincidence as the non-Gaussianity in the local model is produced by the modulation of the small scale 
inflaton power spectrum by the large scale inflaton fluctuations that have already left the horizon and frozen out. 
The parameter $f_{NL}$ is a measure of the non-linear coupling between these different scales. Likewise the large 
scale matter overdensity modulates the small scale Poisson fluctuation power spectrum by altering the local number 
density of radio point sources. In an analogous fashion the bias describes the coupling between the different scales.
As discussed in the previous section any bispectrum present in the data can bias the estimator.

\section{Anisotropy Mechanisms}

Now we will discuss the various physical effects considered in this paper -- radio point source emission, the thermal
Sunyaev-Zeldovich effect and the integrated Sachs-Wolfe effect.

\subsubsection{Radio Point Sources}

The temperature anisotropy induced by unresolved radio point sources can be expressed as an integral over
their flux distribution function
\begin{equation}\label{eq:flux}
\frac{\Delta T}{T}(\hat{\bm n},\nu) = \frac{1}{c_{\nu}}
\int_0^{\bar{S}(\hat{\bm n})} dS S \frac{dN}{dS}(S,\nu;\hat{\bm n}).
\end{equation}
The conversion between the temperature and intensity fluctuations is 
\begin{equation}
c_{\nu} = \frac{\partial B_{\nu}}{\partial \ln{T}}(T_{CMB}), 
\end{equation}
where $T_{CMB} = 2.728$ K and $B_{\nu}(T)$ is a blackbody frequency distribution.

Note that we have allowed both the upper flux limit for unresolved radio point sources and their number density to be spatially 
inhomogeneous. As we will describe below, these spatial inhomogeneities will correlate with other signals present in the data to 
produce bispectra in the observed CMB data.

The actual values for the bispectra calculated in this paper depend on the radio point sources properties (both flux and redshift distributions) 
and on the flux cut at a given frequency for a specific experiment. In order to evaluate the residual 
point source contribution to the total estimated bispectrum in the WMAP--3yrs data, we must consider the technique used by WMAP to 
identify and substract radio point sources \cite{Hinshaw07} and the point source mask applied to the data. The WMAP source 
selection criterion does not correspond to a single flux threshold at a given frequency, rather it requires that a candidate 
source should be seen with a minimal statistical significance in all channels \citep{Pierpa03}.  As Yadav \& Wandelt's 
results \citep{Yadav07}, which are the motivation for this paper, were derived using the V and W frequencies bands, the radio
point source populations at these frequencies is the most relevant. Unfortunately, as most radio sources are stronger at low frequencies, 
they tend to be detected with higher significance in the K -- Q bands than in the V -- W ones. 

In addition from a blind search,  at 20--30 GHz it is possible to use lower--frequency catalogs of point sources as tracers for 
detection \citep{LopezCaniego07}. As a result, the source population at such frequencies is much 
better characterized than the one at higher frequencies. By considering flux number counts at low frequencies it is possible 
to give an estimate of the flux above which the detected point sources create a complete catalog. 
This flux threshold is estimated to be above 1.1 Jy at 23 GHz (K Band) \cite{LopezCaniego07,Gonzalez08}, while in the W band 
(94 GHz) the number counts are not sufficiently well determined to allow such estimate.
An alternative blind search technique applied to the WMAP V and W band data increases the number of sources found in the V band 
by 50\% compared to the WMAP team results \citep{ChenWright07}. These results include some sources that were not contained in 
the WMAP point source mask. The new sources found, however, do not seem to represent a different population than the ones
previously detected. As the WMAP 5 year data is now available, more work on point source characterization at the frequencies 
where the CMB science is derived should be possible. 

The actual residual contribution of point sources depends upon the mask applied to the observed map. In the case of WMAP, this 
mask considers the WMAP detected sources (about 300) as well as some bright sources from other low--frequencies catalogs, for 
a total of approximately seven hundred sources masked. The actual selection function that this procedure imposes at the V and 
W bands is poorly understood, and indeed some sources detected by \cite{ChenWright07} were not masked. However, while the 
detection threshold implied by this whole procedure is poorly defined, there is good agreement in the residual point sources 
power spectrum contribution as derived by different authors \citep{Huffenberger06,ChenWright07,Hinshaw07}. This can be 
approximately translated in a flux threshold of 0.6 Jy in the Q band \citep{Huffenberger06} and 0.75 Jy in the V band 
\cite{ChenWright07}. For illustrative purposes, we will adopt here an approximate estimate for the flux cut-off of 0.7 Jy 
in all WMAP bands.  
In \S \ref{sec:results} we discuss the dependence of our results on this choice.

The Planck satellite will have better resolution and sensitivity, resulting in a lower detection threshold for point sources. 
In the following, we make predictions of the bispectrum expected in the final Planck maps, considering the detection thresholds 
for a 95\% complete sample derived  by \cite{LopezCarniego06} using realistic Planck sky simulations. The flux cut-offs for both
WMAP and Planck, as well as, the adopted instrument noise parameters are given in Table \ref{table:exp_info}.

\begin{table}
\begin{center}
\begin{tabular}{c|c|c|c}
\hline
\hline
Frequency (GHz) & $\bar{S}$ (Jy) & FWHM (arcmin) & $\Delta T/T$ \\
\hline
WMAP -- 33 (Ka) & 0.7 & 41 & 5.7  \\
WMAP -- 41 (Q)  & 0.7 & 28 & 8.2 \\
WMAP -- 61 (V)  & 0.7 & 21 & 11.0  \\
WAMP -- 94 (W)  & 0.7 & 13 & 18.3 \\
\hline
Planck -- 30  & 0.33 & 33 & 1.6 \\
Planck -- 44  & 0.36 & 23 & 2.4  \\
Planck -- 70  & 0.34 & 14 & 3.6 \\
Planck -- 100 & 0.13 & 11 & 1.6  \\
\hline
\hline
\end{tabular}\\
\end{center}
\caption{\label{table:exp_info} Radio point source flux threshold, FWHM, and instrument pixel noise (in $10^{-6}$)
for the relevant WMAP and Planck frequency bands. For WMAP the various band names are also listed.}
\end{table}

Finally, in order to compute the bispectra implied by point sources below a given flux, we adopt the source counts predictions 
of Toffolatti et al \cite{Toffolatti98} rescaled by a factor 0.8, as suggested by the matching of these predictions with the 
actual number counts at fluxes above 1.1 at 41 GHz obtained by \cite{Gonzalez08}. We will use the radio point source bias $b^{PS} 
\simeq 1.7$ \cite{Smith07, Blake04, Boughn02} as inferred for low frequency radio point sources. This value is uncertain and almost
definite varies for the various population type that constitute the high frequency sample.

It is difficult to constrain the redshift distribution due to the lack of optical studies of $60-96 {\rm GHz}$ source population, 
work at $23 {\rm GHz}$ has been conducted by 
Gonzalez et al. for fluxes above 1 Jy \cite{Gonzalez08}. At these fluxes and frequencies the number counts are dominated by 
QSOs for relatively high redshifts. This population is well approximated by the following analytical formula:
\begin{equation}\label{eq:n_PS}
n^{PS}(z)  \propto 0.75 \times e^{-(z-z_0)^2/(2 \sigma)^2},
\end{equation}
with $z_0 = 0.95$ and $\sigma$ is 0.4 (0.9) for $z < 1$ ($z > 1$). This redshift distribution is in agreement with recent 
studies of radio sources populations at ~90 GHz with ATCA \citep{Sadler07}. In addition, there is a small (10--15 \%) of 
the total population that consists of radio loud galaxies which peaks at much lower redshifts ($z \le 0.1$). Given the small 
number of sources, it is difficult to derive an appropriate fitting formula for this other population. In the following, 
we will adopt the following distribution of low redshift sources
\begin{equation}\label{eq:n_PSlow}
n^{PS}(z) \propto 0.25 \times 10^{-3z},
\end{equation}
that provides a better fit to counts found by Gonzalez et al. \cite{Gonzalez08}.
We will take the sum of Eqs.\ref{eq:n_PS} \& \ref{eq:n_PSlow} with a common proportionality factor determined by  requiring 
$n^{PS}(z)$ to integrate to unity over the range $z=0$ to $z=3.1$. The relative amplitude of the QSO and radio galaxy contribution
to the sources is derived from the optical indentifications of Gonzalez et al \cite{Gonzalez08}.
This fit will be called model 1. Gonzalez et al. \cite{Gonzalez08} provide a fit to a theoretical model of the luminosity functions of 
the various source populations as derived from \cite{DeZotti05}. To understand how the uncertainty in the source redshift distribution 
affects our results, we also do calculations with the model of Gonzalez \cite{Gonzalez08}; this will be called model 2.

WMAP only resolves the high-flux sources which are typically dominated by AGN. Most likely lower flux sources consist of a 
different population (e.g. \cite{PierpaPerna04, DeZotti05}) and therefore have a different redshift distribution with possibly 
more weight either at lower or higher redshifts. An analogous redshift analysis on higher frequencies catalog is strongly needed, but 
is not available at this time. Future investigations of radio point sources catalogs plus WMAP and Planck results are higher
frqeuencies will help clarify this issue. For the aims of this paper we take Eq.~\ref{eq:n_PS} to be the redshift distribution at 
all frequencies and fluxes and keep in mind the potential uncertainty that this assumption introduces.

\subsubsection{Thermal Sunyaev-Zeldovich Effect}

The hot plasma in the intra-cluster medium will produce temperature anisotropies via Thomson scattering of the 
incident CMB photons; this is the well-known thermal Sunyaev-Zeldovich (SZ) effect (see \citet{Carlstrom02} for 
a review). Following the model of Komatsu \& Seljak \cite{Komatsu02a,Komatsu01}, the temperature anisotropies 
produced by the SZ effect can be expressed as an integral over the cluster mass distribution function
\begin{equation}\label{eq:sz}
\frac{\Delta T}{T}(\hat{\bm n},\nu) = g_{\nu} \int dz \frac{dV}{dz} \int_0^{\infty} dM y(M,z)
\frac{dn}{dM}[M,z;\hat{\bm n \chi(z)}],
\end{equation}
here the frequency dependence of the thermal SZ effect is given by
\begin{equation}
g_{\nu} = x\frac{e^x+1}{e^x-1} - 4,
\end{equation}
where $x = h\nu/k_B T_{\rm CMB}$. The Compton y-parameter is related to the line-of-sight integral of the the 
cluster's thermal pressure, $dn/dM$ is the cluster mass function and the volume element is
\begin{equation}
\frac{dV}{dz} = \frac{c}{H(z)} \chi^2(z),
\end{equation}
where $\chi(z)$ is comoving distance to redshift z. The details of the implementation of this model are extensively 
discussed in \cite{Komatsu02a}. Ignoring clustering terms in the power spectrum, we find the thermal SZ point source 
power spectrum
\begin{equation}\label{eq:cl_sz}
C^{SZ}_{\ell}(\nu) = g^2_{\nu} \int dz \frac{dV}{dz} \int_0^{\infty} dM y_{\ell}^2(M,z) \frac{dn}{dM}(M,z).
\end{equation}

In a similar fashion to the radio point sources, the number density of massive clusters will be changed by both the
biasing effect of the large scale matter overdensity and gravitational lensing magnification. 
In order to determine how these processes will affect the SZ power spectrum we need to know the redshift distribution
of power in the SZ effect. The weighted redshift distribution of power produced by the SZ effect can be expressed 
as the integral over halo mass of the Sheth-Tormen \citep{Sheth99} halo mass function 
\begin{equation}\label{eq:n_SZ}
n^{SZ}(z) \propto  \frac{dV}{dz} \int_{M_{\rm min}}^{M_{\rm max}} dM M^{2\alpha} \frac{dn}{dM}(M,z).
\end{equation}
The cluster y-parameter -- mass scaling relationship
slope is taken to be $\alpha = 1.6$ \cite{Nagai06}. The limits of integration are taken to be $M_{\rm min}= 10^{14} M_{\odot}$ 
and $M_{\rm max} = 5\times 10^{15} M_{\odot}$. The redshift distribution is normalized so it will integrate to unity. We also 
need the bias weighted redshift distribution of power produced by the SZ effect can be expressed as the integral over the 
Sheth-Tormen halo mass function 
\begin{equation}\label{eq:bn_SZ}
(bn)^{SZ}(z) \propto  \frac{dV}{dz} \int_{M_{\rm min}}^{M_{\rm max}} dM M^{2\alpha} \frac{dn}{dM}(M,z)~b(M,z),
\end{equation}
where $b(M,z)$ is the standard bias of the Sheth-Tormen mass function. 

\subsubsection{Integrated Sachs-Wolfe Effect}

Most of the power in the CMB temperature anisotropies is produced at high redshift during recombination. There will be
very little cross-correlation between these temperature anisotropies and the low redshift matter overdensity responsible
for altering the small scale point source power spectrum. However additional CMB temperature anisotropies can be generated
at low redshift via the Integrated Sachs-Wolfe (ISW) effect if the gravitational potential fluctuations are evolving in time.
In a fully matter dominated regime the gravitational potential fluctuations are static, however as the universe becomes dark
energy dominated the ISW effect can occur.

The temperature anisotropy produced by the ISW effect can be expressed as
\begin{eqnarray}
\frac{\Delta T}{T}(\hat{\bm n}) &=& -2 \int dz \frac{\partial \Phi}{\partial z}, \\
&=& 3 H^2_0 \Omega_M \int dz [(1+z) D(z)]' \int \frac{d^3{\bm k}}{(2\pi)^3} e^{i{\bm k} \cdot \hat{\bm n} \chi} 
\frac{\delta({\bm k})}{k^2},
\end{eqnarray}
where $D(z)$ is the linear theory growth function and the prime denotes differentiation with respect to $z$.

\begin{figure}
\centering
\includegraphics[width=10cm,height=10cm]{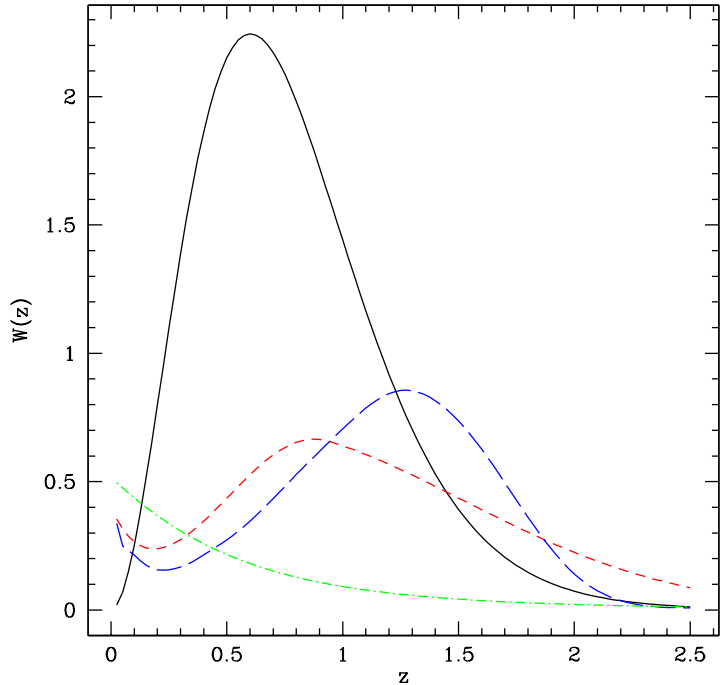}
\caption{\label{fig:nz} Redshift weight functions for the ISW effect (green, dot-dashed); radio point 
sources -- model 1 (red, dashed) and model 2 (blue, long-dashed); and thermal SZ effect (solid, black).}
\end{figure}

In Fig. \ref{fig:nz} we show the redshift weight functions for the ISW effect $[(1+z)D(z)]'$ (green, dot-dashed); the 
radio point sources $b^{PS} n^{PS}(z) D(z)$  -- model 1 (red, dashed) and  model 2 (blue, long-dashed); and the thermal 
SZ effect $(bn)^{SZ}(z) D(z)$ (solid, black). The overlap of these weight fuctions will determine the amplitude of the 
cross-correlation spectra as described in \S \ref{sec:cross} \& \ref{sec:mag} and shown in Figs. \ref{fig:cross_power} 
\& \ref{fig:mag_power}.

\subsection{Point Source Bispectrum}\label{sec:ps_bisp}

The simplest bispectrum form produced in the CMB is due to radio point source Poisson fluctuations. 
The reduced bispectrum is independent of scale and can be written as
\begin{equation}
b_{\ell_1, \ell_2, \ell_3} = c_{\nu}^{-3} \int_0^{\bar{S}} dS S^3 \frac{\bar{dN}}{dS}(S,\nu).
\end{equation}
This bispectrum component has been detected in the WMAP data \cite{Komatsu03}. Since its functional form is not similar 
to the local model it will not significantly bias the estimator and we will ignore it as did Yadav \& Wandelt. The five-year
WMAP analysis \cite{Komatsu08} includes estimates (and corrections) for the estimator bias produced by this bispectrum form.

\subsection{Number Density Modulation}\label{sec:cross}

The Poisson fluctuation power spectrum in a certain region of the sky can be written as integral over the distribution of 
sources in that region. If the anisotropic component of the power spectrum correlates with any other signal present in the 
data, non-Gaussian correlation functions will be generated. In this subsection we will focus on correlation of the large 
scale matter overdensity, which bias the number density of point sources, with the ISW effect.

The power spectrum produced by radio point sources in direction $\hat{\bm n}$ is
\begin{equation}
C^{PS}(\hat{\bm n}) = c_{\nu}^{-2} \int_0^{\bar{S}} dS S^2 \frac{dN}{dS}(S,\nu;\hat{\bm n}),
\end{equation}
where the anisotropic point source distribution can be expressed in terms of the matter overdensity as
\begin{equation}\label{eq:ps_n}
\frac{dN}{dS}(S,\nu;\hat{\bm n}) = \frac{\bar{dN}}{dS}(S,\nu) \left[ 1 + b^{PS}\int dz 
n^{PS}(z) \delta(\hat{\bm n},z) \right],
\end{equation}
the mean point source distribution was described in \S \ref{sec:ps}. The large scale matter overdensity, which biases the 
local number density of point sources as shown in Eq. \ref{eq:ps_n}, will be correlated with the large scale CMB temperature 
anisotropies produced by the ISW effect and the following reduced bispectrum will be induced
\begin{equation}
b_{\ell_1, \ell_2, \ell_3} = 2 \bar{C}^{PS} (X^{PS}_{\ell_1}  + \mathrm{cyc.}).
\end{equation}
Here the isotropic source distribution $\bar{dN}/dS$ leads to an isotropic power spectrum $\bar{C}^{PS}$. The matter-ISW 
cross correlation spectrum can be expressed as
\begin{equation}\label{eq:x_ps}
X^{PS}_{\ell} = \frac{3 H^2_0 \Omega_M b^{PS}}{\ell^2} \int \frac{H(z) dz}{c} P\left[\frac{\ell}{\chi(z)}\right]
 D(z) n^{PS}(z) [(1+z) D(z)]' 
\end{equation}
where $P(k)$ is the matter power spectrum and $b^{PS} \simeq 1.7$ is the radio point source bias. The cross-correlation
spectrum is positive because matter overdensities correspond to potential wells. At low redshift as the amplitude of the 
potential wells decay the CMB photons experience a net blue-shift, thus the cross-correlation is positive. In Eq. \ref{eq:x_ps} 
we have employed Limber's approximation to simplify the cross-correlation spectrum. The function $n^{PS}(z)$ is the redshift 
probability distribution function of the radio point sources defined in Eq. \ref{eq:n_PS}. The large scale matter overdensity 
affects the thermal SZ power spectrum in an analogous fashion. 

\begin{figure}
\centering
\includegraphics[width=10cm,height=10cm]{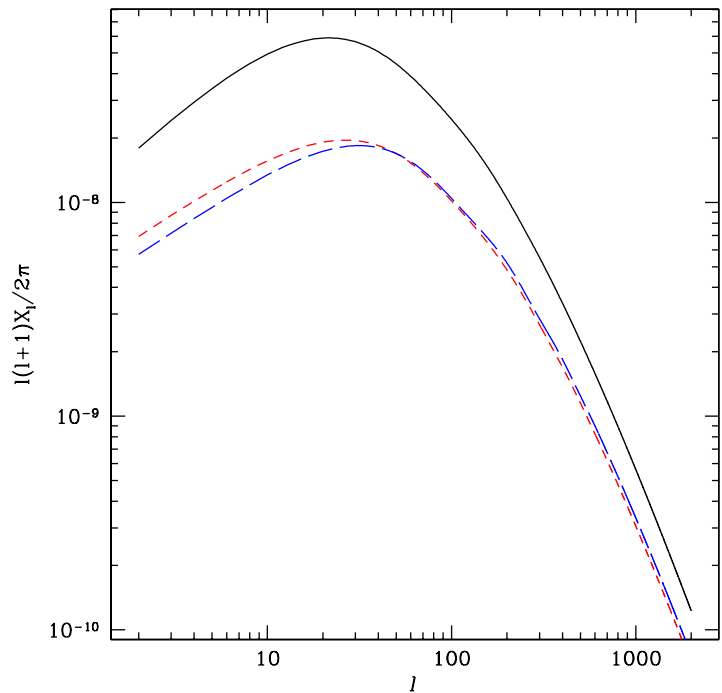}
\caption{\label{fig:cross_power} Density-ISW cross-correlation spectrum for thermal SZ (solid, black); radio point 
sources -- model 1 (red, dashed) and model 2 (blue, long-dashed).}
\end{figure}

In Fig. \ref{fig:cross_power} we show the matter-ISW cross correlation spectrum for the thermal SZ effect (solid, black) and
the radio point sources -- model 1 (red, dashed), model 2 (blue, long-dashed). The thermal SZ effect has the largest cross
correlation because the clusters tend to be located at lower redshift than the radio point sources. The ISW effect primarily
occurs at low redshift once the universe is strongly dark energy dominated, so it most strongly overlaps with the SZ effect.

\begin{figure}
\centering
\includegraphics[width=10cm,height=10cm]{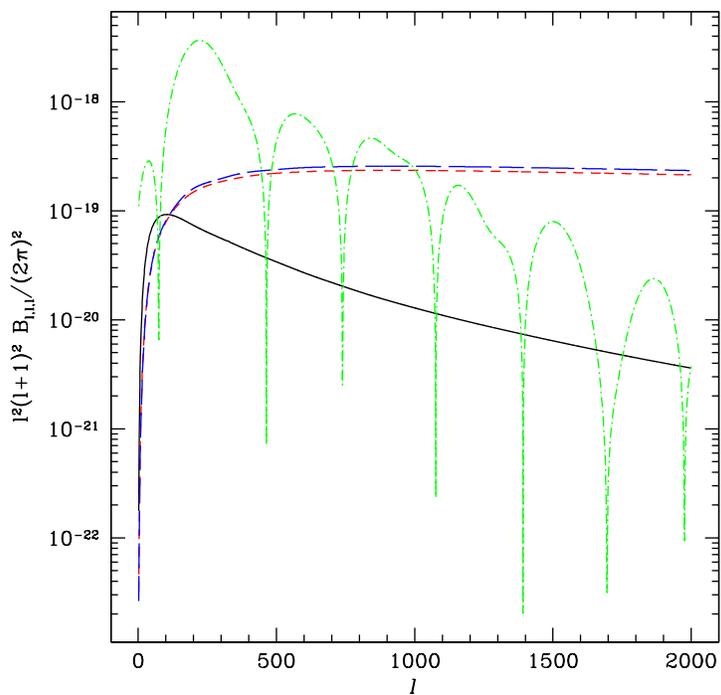}
\caption{\label{fig:cross_bispect} Number density modulation bispectrum for thermal SZ (solid, black); radio point 
sources -- model 1 (red, dashed) and model 2 (blue, long-dashed). The equilateral shape $(\ell_1 = \ell_2 = \ell_3 = \ell)$ 
of the local model ($f_{NL} = 1$) (green, dot-dashed) is also shown for reference.}
\end{figure}

In Fig. \ref{fig:cross_bispect} we show the number density modulation bispectrum for the thermal SZ effect (solid, black) and
the radio point sources -- model 1 (red, dashed), model 2 (blue, long-dashed). Also shown for reference is the equilateral shape 
$(\ell_1 = \ell_2 = \ell_3 = \ell)$ of the primordial local model bispectrum ($f_{NL} = 1$) (green, dot-dashed). The equilateral 
shape of the local model bispectrum changes signs, the zero-crossing are obvious from the plot. This is a consequence of the radiative 
transfer functions producing both hot and cold regions on the sky. The collapsed shape always has the same sign, opposite the sign 
of $f_{NL}$. For the local model the collapsed bispectra have the highest signal-to-noise. 

\subsection{Magnification Modulation}\label{sec:mag}

The distribution of matter along the line-of-sight between the observer and the sources will gravitationally lens these
point sources. The priniciple effect of gravitational lensing will be to change the source density by magnifying
and de-magnifying certain regions of the sky. The magnification in a given direction can be written, to first order, as 
\begin{equation}
\mu(\hat{\bm n}) \simeq 1 + 2 \kappa(\hat{\bm n}). 
\end{equation}
The convergence field distorting a background source at $z$ is related to matter overdensity as
\begin{equation}
\kappa(\hat{\bm n}, z) = \frac{3 \Omega_M H^2_0}{2} \int \frac{c dz'}{H(z')} (1+z') \frac{\chi(z')}{\chi(z)} [\chi(z)-\chi(z')]
\delta(\hat{\bm n},z'),
\end{equation}
and the average magnification of SZ point sources in a given direction is then given by
\begin{equation}
\kappa(\hat{\bm n}) = \int dz n^{SZ}(z) \kappa(\hat{\bm n}, z).
\end{equation}
The observed SZ cluster number density in a given direction will be related to the intrinsic number density as
\begin{eqnarray}
\frac{dn^{obs}}{dM}(M,\hat{\bm n}) &=& \frac{1}{\mu(\hat{\bm n})} \frac{dn}{dM}(M,\hat{\bm n}), 
\\
&\simeq& [1 - 2\kappa(\hat{\bm n})] \frac{dn}{dM}(M,\hat{\bm n}).
\end{eqnarray}
The convergence field is correlated with the CMB temperature anisotropies produced via the ISW effect,
\begin{equation}\label{eq:m_sz}
M^{SZ}_{\ell} = \frac{9 \Omega^2_M H^4_0}{2\ell^2} \int_0^{\infty} dz n^{SZ}(z) 
\int_0^{z} dz_1 P\left[\frac{\ell}{\chi(z_1)}\right] (1+z_1) D(z_1) [(1+z_1) D(z_1)]' \frac{\chi(z_1)}{\chi(z)} [\chi(z)-\chi(z_1)].
\end{equation}
Again we have evaluated the cross-correlation according to Limber's approximation and the convergence-ISW cross-correlation spectrum
is positive. This magnification effect will result in the following reduced bispectrum
\begin{equation}\label{eq:lensing_sz}
b_{\ell_1, \ell_2, \ell_3} = -2 [M^{SZ}_{\ell_1} (C^{SZ}_{\ell_2} + C^{SZ}_{\ell_3}) 
+ M^{SZ}_{\ell_2} (C^{SZ}_{\ell_1} + C^{SZ}_{\ell_3}) 
M^{SZ}_{\ell_3} (C^{SZ}_{\ell_1} + C^{SZ}_{\ell_2})].
\end{equation}
The bispectrum is negative because large scale CMB hot spots (positive ISW effect) correlate with positive magnification which always
reduces the amplitude of point source Poisson fluctuations.

The radio point source selection function is expressed in terms of the observed flux which can be affected by gravitational 
lensing due matter along the line-of-sight. The gravitational lensing magnification will modulate the flux cut-off and be 
correlated with the ISW temperature anisotropies. The point source dilution effect discussed above will also occur. The 
radio point source Poisson fluctuation power spectrum can be expressed as
\begin{equation}
C^{PS}(\hat{\bm n}) = c_{\nu}^{-2} \int_0^{\bar{S}/\mu(\hat{\bm n})} dS S^2 \frac{1}{\mu(\hat{\bm n})} \frac{dN}{dS}(S,\nu).
\end{equation}
Linearizing in the convergence field, we find that the Poisson fluctuation power spectrum becomes anisotropic
\begin{equation}
C^{PS}(\hat{\bm n}) = 2 c_{\nu}^{-2} \kappa(\hat{\bm n}) \left[ \bar{S}^3 \frac{dN}{dS}(\bar{S},\nu) + \bar{C}^{PS} \right].
\end{equation}
This will result in the following reduced bispectrum
\begin{equation}\label{eq:lensing}
b_{\ell_1, \ell_2, \ell_3} =  -4(M^{PS}_{\ell_1} + {\rm cyc.}) \left[ \bar{S}^3 \frac{dN}{dS} (\bar{S}, \nu) + \bar{C}^{PS} \right],
\end{equation}
where $M^{PS}_{\ell}$ is the radio point source version of Eq. \ref{eq:m_sz}.
Note that gravitational lensing affects the radio point source power spectrum by changing both the upper flux cut-off
and the source counts, whereas the SZ power spectrum is only altered by changes in the local cluster counts. If the SZ
clusters are detected with high signal-to-noise and removed from the CMB maps then flux cut-off modulation effect will
also produce an additional bispectrum term.

\begin{figure}
\centering
\includegraphics[width=10cm,height=10cm]{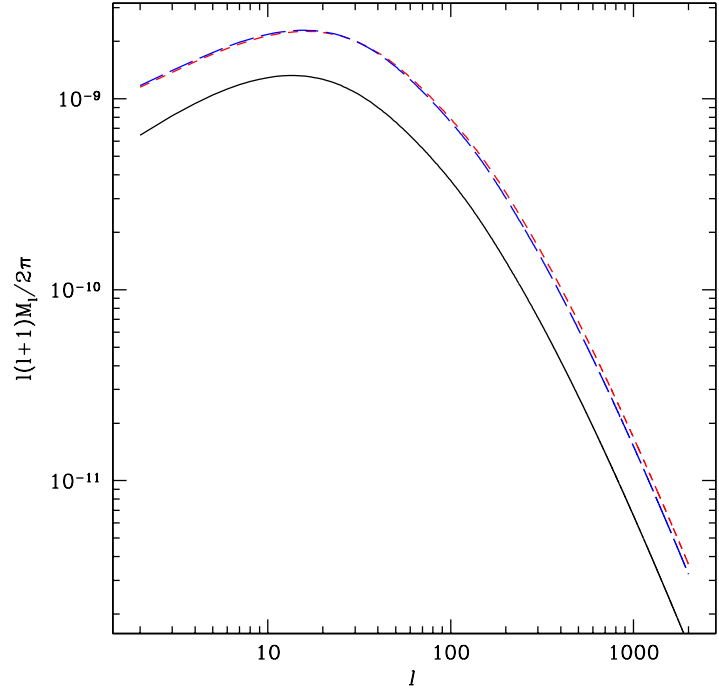}
\caption{\label{fig:mag_power} Convergence-ISW cross-correlation spectrum for thermal SZ (solid, black); radio point 
sources -- model 1 (red, dashed) and model 2 (blue, long dashed)}
\end{figure}

In Fig. \ref{fig:mag_power} we show the convergence-ISW cross correlation spectrum for the thermal SZ effect (solid, black) 
and the radio point sources -- model 1 (red, dashed), model 2 (blue, long-dashed). The model 2 of the radio point sources 
has the largest cross correlation because it predicts that the point sources tend to be located at higher redshift. As 
opposed to the matter-ISW cross-correlation, which requires that the sources lie in the same redshift range over which the 
ISW effect occurs, in this case the point source do not have to be at the same redshift at which the  ISW
occurs  in order for the contribution to be relevant. In fact the point sources 
experience greater magnification if they are at substantially higher redshift than the matter distribution that is 
simultaneously magnifying them and is correlated with the ISW effect. 

\begin{figure}
\centering
\includegraphics[width=10cm,height=10cm]{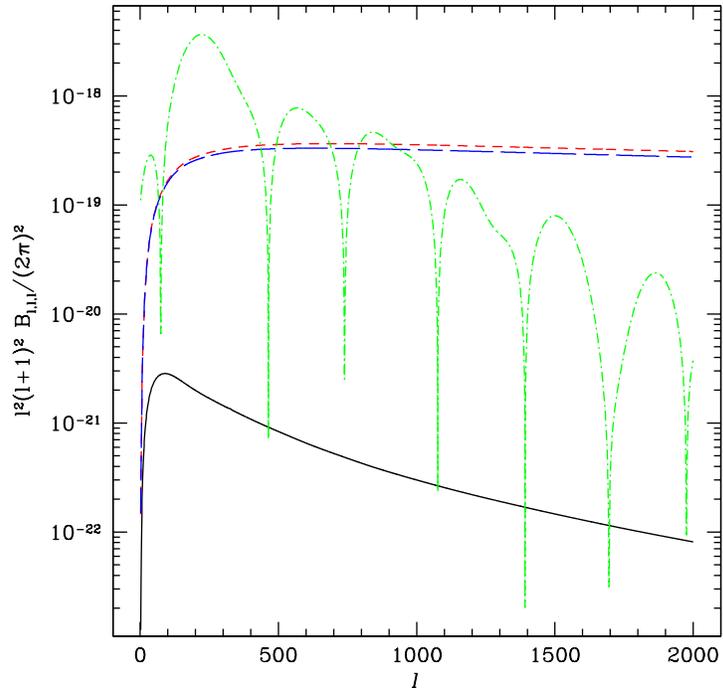}
\caption{\label{fig:mag_bispect} Magnification modulation bispectrum for thermal SZ effect (solid, black); radio point 
sources -- model 1 (red, dashed) and model 2 (blue, long-dashed). The local model bispectrum ($f_{NL} = 1$) --
equilateral shape $(\ell_1 = \ell_2 = \ell_3 = \ell)$ (green, dot-dashed) is also shown for reference.}
\end{figure}

In Fig. \ref{fig:mag_bispect} we show the magnification modulation bispectrum for the thermal SZ effect (solid, black);
radio point sources -- model 1 (red, dashed) and model 2 (blue, long-dashed). The local model bispectrum ($f_{NL} = 1$) 
(green, dot-dashed) is also shown for reference.

\subsection{Selection Modulation}

The amplitude of the Poisson fluctuation power spectrum depends on the number density of radio point sources 
below some flux limit determined by the radio point source removal technique. If the selection criterion used
produces an anisotropic flux limit that is correlated with either the large scale temperature anisotropy
or the instrument noise, then a bispectrum will be produced. If the radio point sources are identified via external 
catalogs, such as NVSS or PMN, then there will be no cross-correlation and therefore no bispectrum. And if
multi-wavelength data is differenced to isolate the power-law frequency dependence of the radio point source, 
then the modulated selection function will produce correlations between the data in different frequency bands.

The simplest method is to removal pixels above some multiple ($\gamma$) of the pixel variance
\begin{equation}
\bar{S}(\hat{\bm n}) = c_{\nu}\left[\gamma \sigma_p - \Delta \Omega \left( \frac{\Delta T}{T}(\hat{\bm n}) 
+ N(\hat{\bm n}) \right)\right]
\end{equation}
The temperature anisotropy and instrument noise have vanishing expectation values, so the mean flux cut-off is
\begin{equation}
\bar{S}_0 = c_{\nu} \gamma \sigma_p,
\end{equation}
where the pixel variance is defined as
\begin{equation}
\sigma^2_p = (\Delta \Omega)^2 \sum_{\ell} \frac{(2\ell + 1)}{4\pi} C^T_{\ell},
\end{equation}
here $C^T_{\ell} = C_{\ell} + C^N_{\ell}$ is the sum of the CMB signal and noise and $\Delta \Omega$ is the pixel 
size. There will be fluctuations about the expected flux cut-off that are correlated with either the temperature 
anisotropy or the instrument noise. Linearizing in these fluctuations, we find the following reduced bispectrum
\begin{equation}
b_{\ell_1, \ell_2, \ell_3} = - \frac{\Delta \Omega}{c_{\nu}} (C^T_{\ell_1} + {\rm cyc.}) \bar{S}^2_0 \frac{dN}{dS}(\bar{S},\nu). 
\end{equation}

The selection criterion used by WMAP is based on an algorithm developed by Tegmark \& de Oliveira-Costa \cite{Tegmark98}.
The map is filtered in order to reduce the importance of the long-wavelength CMB modes. The filtered total temperature
fluctuation in some pixel is
\begin{equation}
y(\hat{\bm n}) = \sum_{\ell m} \frac{1}{C^T_{\ell}} Y_{\ell m}(\hat{\bm n}) (a_{\ell m} + n_{\ell m}) \Delta \Omega 
+ \sum_{\ell} \frac{(2\ell + 1)}{4\pi} \frac{S(\hat{\bm n})}{c_{\nu} C^T_{\ell}}.
\end{equation}
The algorithm removes any pixel that has a value greater than some multiple ($\gamma$) of the filtered map pixel variance
\begin{equation}
\tilde{\sigma}^{-2}_p = (\Delta \Omega)^{-2} \sum_{\ell} \frac{(2\ell + 1)}{4\pi} \frac{1}{C^T_{\ell}}.
\end{equation}
The threshold $\gamma$ is chosen according to some compromise between false positive and negatives.
Since the thresholding is applied to the total signal in a pixel, the corresponding radio point source flux 
cut-off in a pixel is
\begin{equation}
\bar{S}(\hat{\bm n}) = \frac{c_{\nu}}{F} \left[ \gamma \tilde{\sigma}_p 
- \Delta \Omega \sum_{\ell m} \frac{1}{C^T_{\ell}} Y_{\ell m}(\hat{\bm n}) (a_{\ell m} + n_{\ell m}) \right],
\end{equation}
where the normalization is
\begin{equation}
F = \sum_{\ell} \frac{(2\ell + 1)}{4\pi} \frac{1}{C^T_{\ell}}.
\end{equation}
The temperature anisotropy and instrument noise have vanishing expectation values, so the mean flux cut-off is
\begin{equation}
\bar{S}_0 = \frac{c_{\nu}}{F} \gamma \tilde{\sigma}_p.
\end{equation}
There will be fluctuations about this expected value and these fluctuations will be correlated with either the
temperature anisotropy or the instrument noise. Linearizing in these fluctuations, we find the following reduced 
bispectrum
\begin{equation}\label{eq:select}
b_{\ell_1, \ell_2, \ell_3} = - \frac{3 \Delta \Omega}{c_{\nu} F}  \bar{S}^2_0 \frac{dN}{dS} (\nu, \bar{S}_0). 
\end{equation}
In regions with large instrument noise or CMB temperature anisotropy, the radio point source flux cut-off will be 
lowered. This reduces the total number density of radio point sources in that region and therefore the Poisson 
fluctuation power spectrum. This effect explains the negative sign in the reduced bispectrum, Eq. \ref{eq:select}.

Since the filtering applied to the maps produces a bispectrum independent of scale, similar to the point source 
bispectrum described in \S \ref{sec:ps_bisp}, we will ignore it. In reality the actual selection function
is more complicate than this simple filter technique predicts and it might produce a bispectrum of a much different form.
Numerical simulations incorporating the exact selection procedure will need to be done in order to fully determine
its effect on the estimator.

\section{Numerical Results}\label{sec:results}

In this section we will present numerical results for the contamination of the standard non-Gaussianity estimator 
by the various bispectra discussed in this paper. This is done for both the WMAP and Planck instrument noise levels, 
frequency bands and flux cut-offs given in Table \ref{table:exp_info}.

\begin{figure}
\centering
\includegraphics{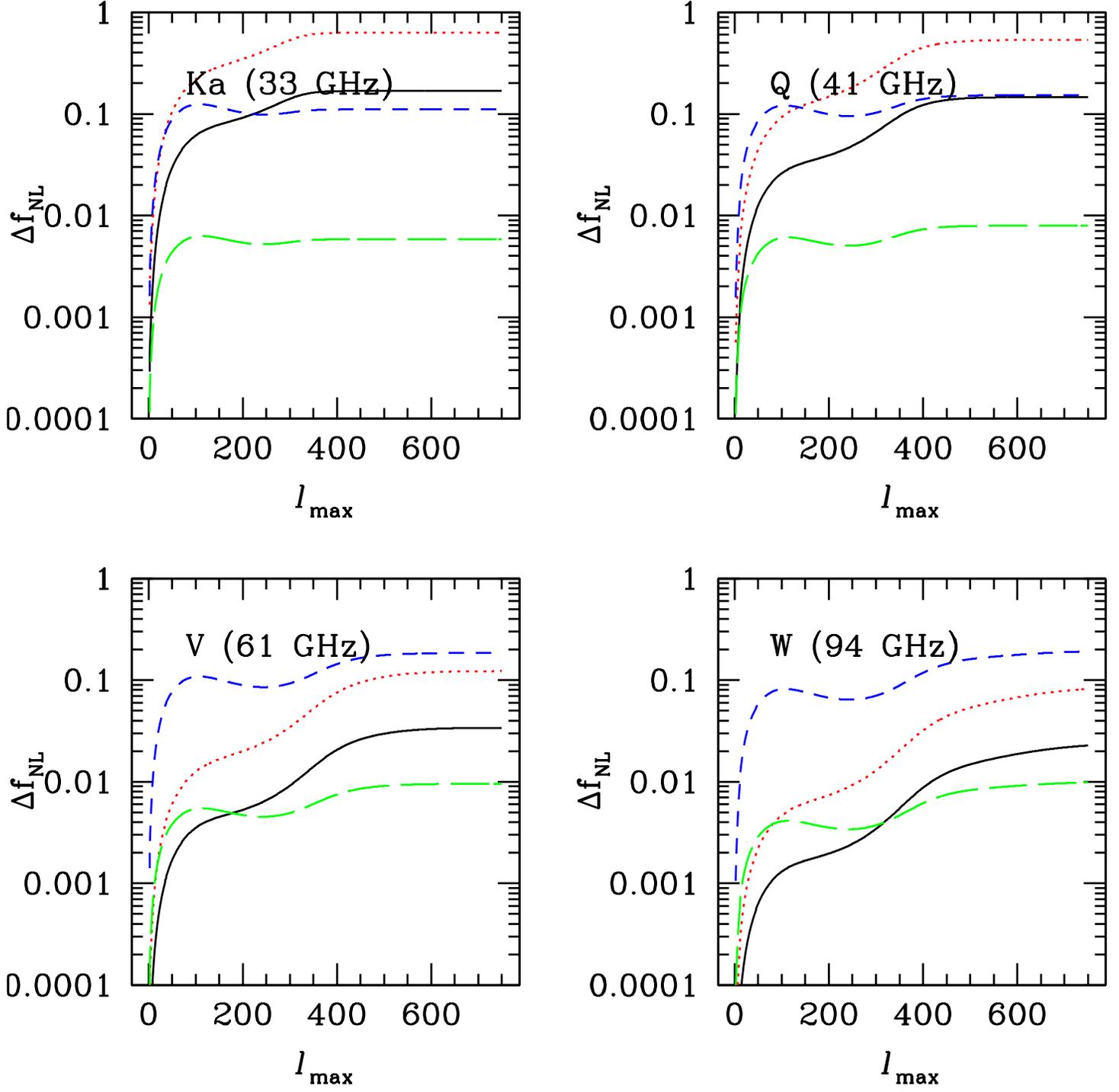}
\caption{\label{fig:wmap} The WMAP estimator bias terms $\Delta f^{\alpha}_{NL}$ as a function of $\ell_{max}$ for radio point 
source density modulation (solid, black); SZ number density modulation (dotted, red); radio point source gravitational lensing 
magnification modulation (dashed, blue); and SZ gravitational lensing magnification modulation (long-dashed, green) in
Ka -- $33~{\rm GHz}$ (upper left); Q -- $40~{\rm GHz}$ (upper right); V -- $61~{\rm GHz}$; and W -- $94~{\rm GHz}$. The density
modulation terms produce a negative bias, while the magnification bias produce a positive bias.}
\end{figure}

In Fig. \ref{fig:wmap} the $f_{NL}$ estimator bias as a function of $\ell_{max}$ is shown for the different bispectra 
-- radio point source number density modulation (solid, black); SZ number density modulation (dotted, red); radio point 
source gravitational lensing magnification modulation (dashed, blue); and SZ gravitational lensing magnification 
modulation (long-dashed, green) with WMAP instrument noise. The bias plots are shown for the four relevant WMAP frequency
bands -- upper left Ka -- $33~{\rm GHz}$; upper right Q -- $40~{\rm GHz}$; lower left V -- $61~{\rm GHz}$; 
lower right W -- $94~{\rm GHz}$. The magnification modulation effect produces a positive bias since its bispectrum is 
negative, while the density modulation effect produces a negative bias. 

Yadav \& Wandelt claim a central value of $f_{NL} = 86.8$ with a standard deviation of $\sigma = 30.0$ for a $2.9 \sigma$
detection of non-Gaussianity \cite{Yadav07}. In the five-year WMAP data the central value is found to be $f_{NL} = 67$ with
a standard deviation of $\sigma = 31$ \cite{Komatsu08}.
At $\ell_{max} = 750$ the total estimator bias $\Delta f_{NL} = 0.35$ in the Ka band, $\Delta f_{NL} = 0.24$ in the Q band,
$\Delta f_{NL} = -0.097$ in the V band and $\Delta f_{NL} = -0.13$ in W band. Since the density modulation and the magnification 
modulation bispectra have different signs they partially cancel and reduce the overall effect. At low frequency the radio point
source magnification modulation bispectrum is the most important so the bias is positive. At higher frequencies the SZ density
modulation bispectrum dominates which makes the bias negative. These numbers should be compared to the estimator bias produced 
by the radio point source Poisson fluctuation bispectrum. Komatsu et al \cite{Komatsu08} have estimated this bias to be 
$\Delta f_{NL} \simeq -(3-5)$ at $\ell_{max} = 700$.

\begin{figure}
\centering
\includegraphics{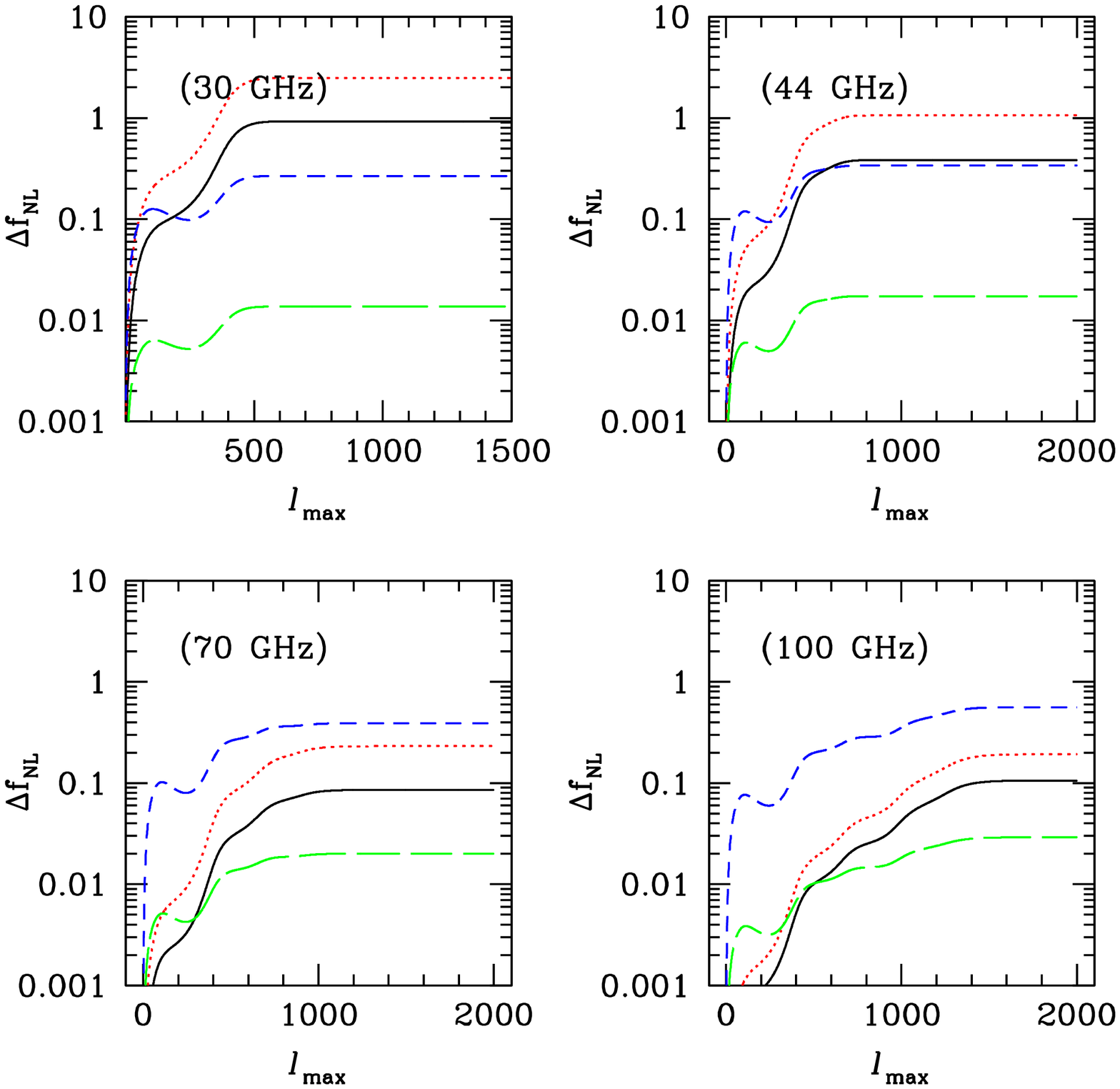}
\caption{\label{fig:planck} The Planck estimator bias terms $\Delta f^{\alpha}_{NL}$ for $30~{\rm GHz}$ (upper left);
$44~{\rm GHz}$ (upper left); $70~{\rm GHz}$ (lower left) and $100~{\rm GHz}$ (lower right). The curves are the same 
as Fig. \ref{fig:wmap}. The density modulation terms produce a negative bias, while the magnification bias produce a positive bias.}
\end{figure}

In Fig. \ref{fig:planck} the $f_{NL}$ estimator bias as a function of $\ell_{max}$ is shown for the different bispectra 
-- radio point source number density modulation (solid, black); SZ number density modulation (dotted, red); radio point 
source gravitational lensing magnification modulation (dashed, blue); and SZ gravitational lensing magnification modulation 
(long-dashed, green) for Planck at $30~{\rm GHz}$ (upper left); $44~{\rm GHz}$ (upper left); $70~{\rm GHz}$ (lower left) 
and $100~{\rm GHz}$ (lower right).

Estimates of the sensitivity on $f_{NL}$ achievable with Planck suggest that $\Delta f_{NL} \simeq 10$ at 95 \% C.L.using just 
temperature information and $\Delta f_{NL} \simeq 5$ at 95 \% C.L also including polarization. Summing the various biases we find
$\Delta f_{NL} = 1.3$ at $\nu = 30$ GHz, $\Delta f_{NL} = 0.34$ at $\nu = 44$ GHz, $\Delta f_{NL} = -0.25$ at $\nu = 70$ GHz
and $\Delta f_{NL} = -0.48$ at $\nu = 100$ GHz. These results imply that a good knowledge of point source
properties is important if Planck will be able to achieve its full potential in constraining primordial non-Gaussianity.

There are uncertainties in the models we have used to describe the radio point sources. These model uncertainities will 
directly lead to uncertainities in the above predictions. As can be directly seen in Figs. \ref{fig:cross_power} -- 
\ref{fig:mag_bispect}, the differences between the radio point source redshift distributions, model 1 and model 2, are 
not significant. This is not surprising as the ISW kernel, with which the radio point source redshift distributions are 
being cross-correlated, is quite broad. The redshift distributions are supposed to trace both the high--flux source 
populations at low frequencies. However, they may not be representative of the source populations at all frequencies 
considered here for lower flux cut thresholds. If a lower flux cut implies a higher population of low--redshift objects, 
the amplitude of the ISW-density cross-correlation spectrum would be increased. The decrease in the flux cut-off will 
decrease the Poisson fluctuation power spectrum, so the change in the bias of the $f_{NL}$ estimator is not clear. We 
also note that reducing the flux cut--off from the WMAP to the Planck level does not reduce the bias implied on $f_{NL}$ 
as the Planck noise levels and beam sizes are also smaller. 

\section{Conclusion}\label{sec:conc}

In this paper we analyzed the effect of point sources, both due to radio emission and the cluster SZ effect, on the 
estimation of primordial non-Gaussianity in the cosmic microwave background. The standard non-Gaussianity estimator is 
sensitive to any bispectrum present in the data. In addition to the standard Poisson fluctuation bispectrum, we found 
that cross-correlations between the radio point source and SZ power spectra and either the CMB temperature anisotropies 
or instrument noise can produce bispectra. These bispectra have forms somewhat similar to the local model, which is the 
standard bispectrum form used to search for primordial non-Gaussianity in CMB data. These similarities are not accidental, 
but occur because the same basic principle generates the non-Gaussianity in these cases. Due to this similarity it will 
be much more difficult to distinguish this non-Gaussianity from the primordial signal than other secondary bispectra with 
shapes which can be quite different.
 
A related paper by Serra \& Cooray \cite{Serra08} has examined different secondary bispectra, but has reached similar conclusions. 
They examined the bispectra produced by the cross-correlation of the thermal SZ effect and the gravitational lensing of the 
primary CMB anisotropies. They concluded that the effects are too small to account for the infered values of $f_{NL}$ from
the WMAP data and will start to become important for the Planck dataset.

The estimator bias that we have calculated is quite small and is not able to explain the results found by Yadav \& Wandelt,
although the bispectra considered in our paper will start to become important for the Planck non-Gaussianity analysis. 
Due to the extremely important nature of the detection of primordial non-Gaussianity in the WMAP data, all possible 
alternative explanations should be considered. 

\begin{acknowledgments}
DB acknowledges financial support from the Betty and Gordon Moore Foundation and would like to thank Sterl Phinney, 
Daisuke Nagai and Eiichiro Komatsu for helpful conversations. EP is and NSF--ADVANCE fellow (AST--0649899) also supported 
by NASA grant NNX07AH59G, Planck subcontract 1290790 and JPL SURP award 1314616.
She would also like to thank Kevin Huffenberger and Joachin Gonzales-Nuevo for useful conversations.
We also thank Kendrick Smith for helpful comments on a draft of the paper.
\end{acknowledgments}


\begin{thebibliography}{27}
\expandafter\ifx\csname natexlab\endcsname\relax\def\natexlab#1{#1}\fi
\expandafter\ifx\csname bibnamefont\endcsname\relax
  \def\bibnamefont#1{#1}\fi
\expandafter\ifx\csname bibfnamefont\endcsname\relax
  \def\bibfnamefont#1{#1}\fi
\expandafter\ifx\csname citenamefont\endcsname\relax
  \def\citenamefont#1{#1}\fi
\expandafter\ifx\csname url\endcsname\relax
  \def\url#1{\texttt{#1}}\fi
\expandafter\ifx\csname urlprefix\endcsname\relax\def\urlprefix{URL }\fi
\providecommand{\bibinfo}[2]{#2}
\providecommand{\eprint}[2][]{\url{#2}}

\bibitem[{\citenamefont{{Yadav} and {Wandelt}}(2007)}]{Yadav07}
\bibinfo{author}{\bibfnamefont{A.~P.~S.} \bibnamefont{{Yadav}}}
  \bibnamefont{and} \bibinfo{author}{\bibfnamefont{B.~D.}
  \bibnamefont{{Wandelt}}}, \bibinfo{journal}{ArXiv e-prints}
  \textbf{\bibinfo{volume}{712}} (\bibinfo{year}{2007}), \eprint{0712.1148}.

\bibitem[{\citenamefont{{Spergel} et~al.}(2007)\citenamefont{{Spergel}, {Bean},
  {Dor{\'e}}, {Nolta}, {Bennett}, {Dunkley}, {Hinshaw}, {Jarosik}, {Komatsu},
  {Page} et~al.}}]{Spergel07}
\bibinfo{author}{\bibfnamefont{D.~N.} \bibnamefont{{Spergel}}},
  \bibinfo{author}{\bibfnamefont{R.}~\bibnamefont{{Bean}}},
  \bibinfo{author}{\bibfnamefont{O.}~\bibnamefont{{Dor{\'e}}}},
  \bibinfo{author}{\bibfnamefont{M.~R.} \bibnamefont{{Nolta}}},
  \bibinfo{author}{\bibfnamefont{C.~L.} \bibnamefont{{Bennett}}},
  \bibinfo{author}{\bibfnamefont{J.}~\bibnamefont{{Dunkley}}},
  \bibinfo{author}{\bibfnamefont{G.}~\bibnamefont{{Hinshaw}}},
  \bibinfo{author}{\bibfnamefont{N.}~\bibnamefont{{Jarosik}}},
  \bibinfo{author}{\bibfnamefont{E.}~\bibnamefont{{Komatsu}}},
  \bibinfo{author}{\bibfnamefont{L.}~\bibnamefont{{Page}}},
  \bibnamefont{et~al.}, \bibinfo{journal}{ApJS} \textbf{\bibinfo{volume}{170}},
  \bibinfo{pages}{377} (\bibinfo{year}{2007}), \eprint{arXiv:astro-ph/0603449}.

\bibitem[{\citenamefont{{Maldacena}}(2003)}]{Maldacena03}
\bibinfo{author}{\bibfnamefont{J.}~\bibnamefont{{Maldacena}}},
  \bibinfo{journal}{Journal of High Energy Physics}
  \textbf{\bibinfo{volume}{5}}, \bibinfo{pages}{13} (\bibinfo{year}{2003}),
  \eprint{arXiv:astro-ph/0210603}.

\bibitem[{\citenamefont{{Komatsu}}(2002)}]{Komatsu02}
\bibinfo{author}{\bibfnamefont{E.}~\bibnamefont{{Komatsu}}},
  \bibinfo{journal}{ArXiv Astrophysics e-prints}  (\bibinfo{year}{2002}),
  \eprint{astro-ph/0206039}.

\bibitem[{\citenamefont{{Babich}}(2005)}]{Babich05}
\bibinfo{author}{\bibfnamefont{D.}~\bibnamefont{{Babich}}},
  \bibinfo{journal}{\prd} \textbf{\bibinfo{volume}{72}},
  \bibinfo{pages}{043003} (\bibinfo{year}{2005}),
  \eprint{arXiv:astro-ph/0503375}.

\bibitem[{\citenamefont{{Creminelli} et~al.}(2006)\citenamefont{{Creminelli},
  {Nicolis}, {Senatore}, {Tegmark}, and {Zaldarriaga}}}]{Creminelli06}
\bibinfo{author}{\bibfnamefont{P.}~\bibnamefont{{Creminelli}}},
  \bibinfo{author}{\bibfnamefont{A.}~\bibnamefont{{Nicolis}}},
  \bibinfo{author}{\bibfnamefont{L.}~\bibnamefont{{Senatore}}},
  \bibinfo{author}{\bibfnamefont{M.}~\bibnamefont{{Tegmark}}},
  \bibnamefont{and}
  \bibinfo{author}{\bibfnamefont{M.}~\bibnamefont{{Zaldarriaga}}},
  \bibinfo{journal}{Journal of Cosmology and Astro-Particle Physics}
  \textbf{\bibinfo{volume}{5}}, \bibinfo{pages}{4} (\bibinfo{year}{2006}),
  \eprint{arXiv:astro-ph/0509029}.

\bibitem[{\citenamefont{{Hinshaw} et~al.}(2007)\citenamefont{{Hinshaw},
  {Nolta}, {Bennett}, {Bean}, {Dor{\'e}}, {Greason}, {Halpern}, {Hill},
  {Jarosik}, {Kogut} et~al.}}]{Hinshaw07}
\bibinfo{author}{\bibfnamefont{G.}~\bibnamefont{{Hinshaw}}},
  \bibinfo{author}{\bibfnamefont{M.~R.} \bibnamefont{{Nolta}}},
  \bibinfo{author}{\bibfnamefont{C.~L.} \bibnamefont{{Bennett}}},
  \bibinfo{author}{\bibfnamefont{R.}~\bibnamefont{{Bean}}},
  \bibinfo{author}{\bibfnamefont{O.}~\bibnamefont{{Dor{\'e}}}},
  \bibinfo{author}{\bibfnamefont{M.~R.} \bibnamefont{{Greason}}},
  \bibinfo{author}{\bibfnamefont{M.}~\bibnamefont{{Halpern}}},
  \bibinfo{author}{\bibfnamefont{R.~S.} \bibnamefont{{Hill}}},
  \bibinfo{author}{\bibfnamefont{N.}~\bibnamefont{{Jarosik}}},
  \bibinfo{author}{\bibfnamefont{A.}~\bibnamefont{{Kogut}}},
  \bibnamefont{et~al.}, \bibinfo{journal}{ApJS} \textbf{\bibinfo{volume}{170}},
  \bibinfo{pages}{288} (\bibinfo{year}{2007}), \eprint{arXiv:astro-ph/0603451}.

\bibitem[{\citenamefont{{Pierpaoli}}(2003)}]{Pierpa03}
\bibinfo{author}{\bibfnamefont{E.}~\bibnamefont{{Pierpaoli}}},
  \bibinfo{journal}{\apj} \textbf{\bibinfo{volume}{589}}, \bibinfo{pages}{58}
  (\bibinfo{year}{2003}), \eprint{arXiv:astro-ph/0301563}.

\bibitem[{\citenamefont{{L{\'o}pez-Caniego}
  et~al.}(2007)\citenamefont{{L{\'o}pez-Caniego}, {Gonz{\'a}lez-Nuevo},
  {Herranz}, {Massardi}, {Sanz}, {De Zotti}, {Toffolatti}, and
  {Arg{\"u}eso}}}]{LopezCaniego07}
\bibinfo{author}{\bibfnamefont{M.}~\bibnamefont{{L{\'o}pez-Caniego}}},
  \bibinfo{author}{\bibfnamefont{J.}~\bibnamefont{{Gonz{\'a}lez-Nuevo}}},
  \bibinfo{author}{\bibfnamefont{D.}~\bibnamefont{{Herranz}}},
  \bibinfo{author}{\bibfnamefont{M.}~\bibnamefont{{Massardi}}},
  \bibinfo{author}{\bibfnamefont{J.~L.} \bibnamefont{{Sanz}}},
  \bibinfo{author}{\bibfnamefont{G.}~\bibnamefont{{De Zotti}}},
  \bibinfo{author}{\bibfnamefont{L.}~\bibnamefont{{Toffolatti}}},
  \bibnamefont{and}
  \bibinfo{author}{\bibfnamefont{F.}~\bibnamefont{{Arg{\"u}eso}}},
  \bibinfo{journal}{ApJS} \textbf{\bibinfo{volume}{170}}, \bibinfo{pages}{108}
  (\bibinfo{year}{2007}), \eprint{arXiv:astro-ph/0701473}.

\bibitem[{\citenamefont{{Gonz{\'a}lez-Nuevo}
  et~al.}(2008)\citenamefont{{Gonz{\'a}lez-Nuevo}, {Massardi}, {Arg{\"u}eso},
  {Herranz}, {Toffolatti}, {Sanz}, {L{\'o}pez-Caniego}, and {de
  Zotti}}}]{Gonzalez08}
\bibinfo{author}{\bibfnamefont{J.}~\bibnamefont{{Gonz{\'a}lez-Nuevo}}},
  \bibinfo{author}{\bibfnamefont{M.}~\bibnamefont{{Massardi}}},
  \bibinfo{author}{\bibfnamefont{F.}~\bibnamefont{{Arg{\"u}eso}}},
  \bibinfo{author}{\bibfnamefont{D.}~\bibnamefont{{Herranz}}},
  \bibinfo{author}{\bibfnamefont{L.}~\bibnamefont{{Toffolatti}}},
  \bibinfo{author}{\bibfnamefont{J.~L.} \bibnamefont{{Sanz}}},
  \bibinfo{author}{\bibfnamefont{M.}~\bibnamefont{{L{\'o}pez-Caniego}}},
  \bibnamefont{and} \bibinfo{author}{\bibfnamefont{G.}~\bibnamefont{{de
  Zotti}}}, \bibinfo{journal}{MNRAS} pp. \bibinfo{pages}{48--+}
  (\bibinfo{year}{2008}), \eprint{arXiv:0711.2631}.

\bibitem[{\citenamefont{{Chen} and {Wright}}(2007)}]{ChenWright07}
\bibinfo{author}{\bibfnamefont{X.}~\bibnamefont{{Chen}}} \bibnamefont{and}
  \bibinfo{author}{\bibfnamefont{E.~L.} \bibnamefont{{Wright}}},
  \bibinfo{journal}{ArXiv e-prints} \textbf{\bibinfo{volume}{712}}
  (\bibinfo{year}{2007}), \eprint{0712.3594}.

\bibitem[{\citenamefont{{Huffenberger}
  et~al.}(2006)\citenamefont{{Huffenberger}, {Eriksen}, and
  {Hansen}}}]{Huffenberger06}
\bibinfo{author}{\bibfnamefont{K.~M.} \bibnamefont{{Huffenberger}}},
  \bibinfo{author}{\bibfnamefont{H.~K.} \bibnamefont{{Eriksen}}},
  \bibnamefont{and} \bibinfo{author}{\bibfnamefont{F.~K.}
  \bibnamefont{{Hansen}}}, \bibinfo{journal}{ApJL}
  \textbf{\bibinfo{volume}{651}}, \bibinfo{pages}{L81} (\bibinfo{year}{2006}),
  \eprint{arXiv:astro-ph/0606538}.

\bibitem[{\citenamefont{{L{\'o}pez-Caniego}
  et~al.}(2006)\citenamefont{{L{\'o}pez-Caniego}, {Herranz},
  {Gonz{\'a}lez-Nuevo}, {Sanz}, {Barreiro}, {Vielva}, {Arg{\"u}eso}, and
  {Toffolatti}}}]{LopezCarniego06}
\bibinfo{author}{\bibfnamefont{M.}~\bibnamefont{{L{\'o}pez-Caniego}}},
  \bibinfo{author}{\bibfnamefont{D.}~\bibnamefont{{Herranz}}},
  \bibinfo{author}{\bibfnamefont{J.}~\bibnamefont{{Gonz{\'a}lez-Nuevo}}},
  \bibinfo{author}{\bibfnamefont{J.~L.} \bibnamefont{{Sanz}}},
  \bibinfo{author}{\bibfnamefont{R.~B.} \bibnamefont{{Barreiro}}},
  \bibinfo{author}{\bibfnamefont{P.}~\bibnamefont{{Vielva}}},
  \bibinfo{author}{\bibfnamefont{F.}~\bibnamefont{{Arg{\"u}eso}}},
  \bibnamefont{and}
  \bibinfo{author}{\bibfnamefont{L.}~\bibnamefont{{Toffolatti}}},
  \bibinfo{journal}{MNRAS} \textbf{\bibinfo{volume}{370}},
  \bibinfo{pages}{2047} (\bibinfo{year}{2006}),
  \eprint{arXiv:astro-ph/0606199}.

\bibitem[{\citenamefont{{Toffolatti} et~al.}(1998)\citenamefont{{Toffolatti},
  {Argueso Gomez}, {de Zotti}, {Mazzei}, {Franceschini}, {Danese}, and
  {Burigana}}}]{Toffolatti98}
\bibinfo{author}{\bibfnamefont{L.}~\bibnamefont{{Toffolatti}}},
  \bibinfo{author}{\bibfnamefont{F.}~\bibnamefont{{Argueso Gomez}}},
  \bibinfo{author}{\bibfnamefont{G.}~\bibnamefont{{de Zotti}}},
  \bibinfo{author}{\bibfnamefont{P.}~\bibnamefont{{Mazzei}}},
  \bibinfo{author}{\bibfnamefont{A.}~\bibnamefont{{Franceschini}}},
  \bibinfo{author}{\bibfnamefont{L.}~\bibnamefont{{Danese}}}, \bibnamefont{and}
  \bibinfo{author}{\bibfnamefont{C.}~\bibnamefont{{Burigana}}},
  \bibinfo{journal}{MNRAS} \textbf{\bibinfo{volume}{297}}, \bibinfo{pages}{117}
  (\bibinfo{year}{1998}), \eprint{arXiv:astro-ph/9711085}.

\bibitem[{\citenamefont{{Smith} et~al.}(2007)\citenamefont{{Smith}, {Zahn}, and
  {Dor{\'e}}}}]{Smith07}
\bibinfo{author}{\bibfnamefont{K.~M.} \bibnamefont{{Smith}}},
  \bibinfo{author}{\bibfnamefont{O.}~\bibnamefont{{Zahn}}}, \bibnamefont{and}
  \bibinfo{author}{\bibfnamefont{O.}~\bibnamefont{{Dor{\'e}}}},
  \bibinfo{journal}{\prd} \textbf{\bibinfo{volume}{76}},
  \bibinfo{pages}{043510} (\bibinfo{year}{2007}), \eprint{arXiv:0705.3980}.

\bibitem[{\citenamefont{{Blake} et~al.}(2004)\citenamefont{{Blake}, {Ferreira},
  and {Borrill}}}]{Blake04}
\bibinfo{author}{\bibfnamefont{C.}~\bibnamefont{{Blake}}},
  \bibinfo{author}{\bibfnamefont{P.~G.} \bibnamefont{{Ferreira}}},
  \bibnamefont{and}
  \bibinfo{author}{\bibfnamefont{J.}~\bibnamefont{{Borrill}}},
  \bibinfo{journal}{MNRAS} \textbf{\bibinfo{volume}{351}}, \bibinfo{pages}{923}
  (\bibinfo{year}{2004}), \eprint{arXiv:astro-ph/0404085}.

\bibitem[{\citenamefont{{Boughn} and {Crittenden}}(2002)}]{Boughn02}
\bibinfo{author}{\bibfnamefont{S.~P.} \bibnamefont{{Boughn}}} \bibnamefont{and}
  \bibinfo{author}{\bibfnamefont{R.~G.} \bibnamefont{{Crittenden}}},
  \bibinfo{journal}{Physical Review Letters} \textbf{\bibinfo{volume}{88}},
  \bibinfo{pages}{021302} (\bibinfo{year}{2002}),
  \eprint{arXiv:astro-ph/0111281}.

\bibitem[{\citenamefont{{Sadler} et~al.}(2007)\citenamefont{{Sadler}, {Ricci},
  {Ekers}, {Sault}, {Jackson}, and {De Zotti}}}]{Sadler07}
\bibinfo{author}{\bibfnamefont{E.~M.} \bibnamefont{{Sadler}}},
  \bibinfo{author}{\bibfnamefont{R.}~\bibnamefont{{Ricci}}},
  \bibinfo{author}{\bibfnamefont{R.~D.} \bibnamefont{{Ekers}}},
  \bibinfo{author}{\bibfnamefont{R.~J.} \bibnamefont{{Sault}}},
  \bibinfo{author}{\bibfnamefont{C.~A.} \bibnamefont{{Jackson}}},
  \bibnamefont{and} \bibinfo{author}{\bibfnamefont{G.}~\bibnamefont{{De
  Zotti}}}, \bibinfo{journal}{ArXiv e-prints} \textbf{\bibinfo{volume}{709}}
  (\bibinfo{year}{2007}), \eprint{0709.3563}.

\bibitem[{\citenamefont{{Pierpaoli} and {Perna}}(2004)}]{PierpaPerna04}
\bibinfo{author}{\bibfnamefont{E.}~\bibnamefont{{Pierpaoli}}} \bibnamefont{and}
  \bibinfo{author}{\bibfnamefont{R.}~\bibnamefont{{Perna}}},
  \bibinfo{journal}{MNRAS} \textbf{\bibinfo{volume}{354}},
  \bibinfo{pages}{1005} (\bibinfo{year}{2004}),
  \eprint{arXiv:astro-ph/0407561}.

\bibitem[{\citenamefont{{de Zotti} et~al.}(2005)\citenamefont{{de Zotti},
  {Ricci}, {Mesa}, {Silva}, {Mazzotta}, {Toffolatti}, and
  {Gonz{\'a}lez-Nuevo}}}]{DeZotti05}
\bibinfo{author}{\bibfnamefont{G.}~\bibnamefont{{de Zotti}}},
  \bibinfo{author}{\bibfnamefont{R.}~\bibnamefont{{Ricci}}},
  \bibinfo{author}{\bibfnamefont{D.}~\bibnamefont{{Mesa}}},
  \bibinfo{author}{\bibfnamefont{L.}~\bibnamefont{{Silva}}},
  \bibinfo{author}{\bibfnamefont{P.}~\bibnamefont{{Mazzotta}}},
  \bibinfo{author}{\bibfnamefont{L.}~\bibnamefont{{Toffolatti}}},
  \bibnamefont{and}
  \bibinfo{author}{\bibfnamefont{J.}~\bibnamefont{{Gonz{\'a}lez-Nuevo}}},
  \bibinfo{journal}{AA} \textbf{\bibinfo{volume}{431}}, \bibinfo{pages}{893}
  (\bibinfo{year}{2005}), \eprint{arXiv:astro-ph/0410709}.

\bibitem[{\citenamefont{{Carlstrom} et~al.}(2002)\citenamefont{{Carlstrom},
  {Holder}, and {Reese}}}]{Carlstrom02}
\bibinfo{author}{\bibfnamefont{J.~E.} \bibnamefont{{Carlstrom}}},
  \bibinfo{author}{\bibfnamefont{G.~P.} \bibnamefont{{Holder}}},
  \bibnamefont{and} \bibinfo{author}{\bibfnamefont{E.~D.}
  \bibnamefont{{Reese}}}, \bibinfo{journal}{ARAA}
  \textbf{\bibinfo{volume}{40}}, \bibinfo{pages}{643} (\bibinfo{year}{2002}),
  \eprint{arXiv:astro-ph/0208192}.

\bibitem[{\citenamefont{{Komatsu} and {Seljak}}(2002)}]{Komatsu02a}
\bibinfo{author}{\bibfnamefont{E.}~\bibnamefont{{Komatsu}}} \bibnamefont{and}
  \bibinfo{author}{\bibfnamefont{U.}~\bibnamefont{{Seljak}}},
  \bibinfo{journal}{MNRAS} \textbf{\bibinfo{volume}{336}},
  \bibinfo{pages}{1256} (\bibinfo{year}{2002}),
  \eprint{arXiv:astro-ph/0205468}.

\bibitem[{\citenamefont{{Komatsu} and {Seljak}}(2001)}]{Komatsu01}
\bibinfo{author}{\bibfnamefont{E.}~\bibnamefont{{Komatsu}}} \bibnamefont{and}
  \bibinfo{author}{\bibfnamefont{U.}~\bibnamefont{{Seljak}}},
  \bibinfo{journal}{MNRAS} \textbf{\bibinfo{volume}{327}},
  \bibinfo{pages}{1353} (\bibinfo{year}{2001}),
  \eprint{arXiv:astro-ph/0106151}.

\bibitem[{\citenamefont{{Sheth} and {Tormen}}(1999)}]{Sheth99}
\bibinfo{author}{\bibfnamefont{R.~K.} \bibnamefont{{Sheth}}} \bibnamefont{and}
  \bibinfo{author}{\bibfnamefont{G.}~\bibnamefont{{Tormen}}},
  \bibinfo{journal}{MNRAS} \textbf{\bibinfo{volume}{308}}, \bibinfo{pages}{119}
  (\bibinfo{year}{1999}), \eprint{arXiv:astro-ph/9901122}.

\bibitem[{\citenamefont{{Nagai}}(2006)}]{Nagai06}
\bibinfo{author}{\bibfnamefont{D.}~\bibnamefont{{Nagai}}},
  \bibinfo{journal}{\apj} \textbf{\bibinfo{volume}{650}}, \bibinfo{pages}{538}
  (\bibinfo{year}{2006}), \eprint{arXiv:astro-ph/0512208}.

\bibitem[{\citenamefont{{Komatsu} et~al.}(2003)\citenamefont{{Komatsu},
  {Kogut}, {Nolta}, {Bennett}, {Halpern}, {Hinshaw}, {Jarosik}, {Limon},
  {Meyer}, {Page} et~al.}}]{Komatsu03}
\bibinfo{author}{\bibfnamefont{E.}~\bibnamefont{{Komatsu}}},
  \bibinfo{author}{\bibfnamefont{A.}~\bibnamefont{{Kogut}}},
  \bibinfo{author}{\bibfnamefont{M.~R.} \bibnamefont{{Nolta}}},
  \bibinfo{author}{\bibfnamefont{C.~L.} \bibnamefont{{Bennett}}},
  \bibinfo{author}{\bibfnamefont{M.}~\bibnamefont{{Halpern}}},
  \bibinfo{author}{\bibfnamefont{G.}~\bibnamefont{{Hinshaw}}},
  \bibinfo{author}{\bibfnamefont{N.}~\bibnamefont{{Jarosik}}},
  \bibinfo{author}{\bibfnamefont{M.}~\bibnamefont{{Limon}}},
  \bibinfo{author}{\bibfnamefont{S.~S.} \bibnamefont{{Meyer}}},
  \bibinfo{author}{\bibfnamefont{L.}~\bibnamefont{{Page}}},
  \bibnamefont{et~al.}, \bibinfo{journal}{ApJS} \textbf{\bibinfo{volume}{148}},
  \bibinfo{pages}{119} (\bibinfo{year}{2003}), \eprint{arXiv:astro-ph/0302223}.

\bibitem[{\citenamefont{{Tegmark} and {de Oliveira-Costa}}(1998)}]{Tegmark98}
\bibinfo{author}{\bibfnamefont{M.}~\bibnamefont{{Tegmark}}} \bibnamefont{and}
  \bibinfo{author}{\bibfnamefont{A.}~\bibnamefont{{de Oliveira-Costa}}},
  \bibinfo{journal}{ApJL} \textbf{\bibinfo{volume}{500}}, \bibinfo{pages}{L83+}
  (\bibinfo{year}{1998}), \eprint{arXiv:astro-ph/9802123}.

\bibitem[{\citenamefont{{Serra} and {Cooray}}(2008)}]{Serra08}
\bibinfo{author}{\bibfnamefont{P.} \bibnamefont{{Serra}}} \bibnamefont{and}
  \bibinfo{author}{\bibfnamefont{A.}~\bibnamefont{{Cooray}}},
  \bibinfo{journal}{.} \textbf{\bibinfo{volume}{.}}, \bibinfo{pages}{.}
  (\bibinfo{year}{2008}), \eprint{arXiv0801.3276}.

\bibitem[{\citenamefont{{Komatsu} et~al.}(2008)\citenamefont{{Komatsu},
  {Dunkley}, {Nolta}, {Bennett}, {Gold}, {Hinshaw}, {Jarosik}, {Limon},
  {Larson},  et~al.}}]{Komatsu08}
\bibinfo{author}{\bibfnamefont{E.}~\bibnamefont{{Komatsu}}},
  \bibinfo{author}{\bibfnamefont{J.}~\bibnamefont{{Dunkely}}},
  \bibinfo{author}{\bibfnamefont{M.~R.} \bibnamefont{{Nolta}}},
  \bibinfo{author}{\bibfnamefont{C.~L.} \bibnamefont{{Bennett}}},
  \bibinfo{author}{\bibfnamefont{M.}~\bibnamefont{{Gold}}},
  \bibinfo{author}{\bibfnamefont{G.}~\bibnamefont{{Hinshaw}}},
  \bibinfo{author}{\bibfnamefont{N.}~\bibnamefont{{Jarosik}}},
  \bibinfo{author}{\bibfnamefont{M.}~\bibnamefont{{Limon}}},
  \bibinfo{author}{\bibfnamefont{D.} \bibnamefont{{Larson}}},
  \bibnamefont{et~al.}, \bibinfo{journal}{ApJS} \textbf{\bibinfo{volume}{.}},
  \bibinfo{pages}{.} (\bibinfo{year}{2000}), \eprint{arXiv:astro-ph/0803.0547}.



\end{thebibliography}

\end{document}